\documentclass[12pt]{article}
\date{}
\begin{document}
\title{Can massless neutrinos oscillate in presence of matter?}
\author{{\bf Indranath Bhattacharyya}\vspace{0.2cm}\\
Department of Mathematics\\
Acharya Prafulla Chandra Roy Government College\\Himachal Vihar, Matigara, Siliguri-734010, West Bengal
\\E-mail :
$i_{-}bhattacharyya@hotmail.com$\\} \vskip .1in \maketitle
\baselineskip .3in \noindent
\begin{abstract}
In presence of matter the possibility of flavor oscillation of
massless neutrino is explored. A comparison between vacuum
oscillation and the matter induced oscillation of massless
neutrinos is carried out to examine whether the flavor transition
is possible in the framework of standard model if there is
background matter. The Stodolsky type of equation
describing the neutrino oscillation phenomenon as the motion of a
spherical pendulum in flavor space is deduced. That pendular model
is studied with zero vacuum oscillation frequency implying the
zero neutrino mass evolved in the framework of standard model, but
non-zero frequency arising due to the MSW effect. The implication
of the non-zero term present in the Stodolsky equation at zero
vacuum frequency is addressed properly.
\vspace{0.5cm}
\vspace{0.1cm}
\end{abstract}
\section{Introduction:}
According to the standard model of electroweak theory \cite{Salam}
the neutrino is considered as massless, although a bare minimum
extension of this model can generate the neutrino mass, however,
it is still questionable why the neutrino mass is so small. In
1957 Pontecorvo \cite{Pontecorvo} proposed the concept of neutrino
oscillation. That is nothing but a quantum mechanical phenomenon
in which one type of neutrino flavor can later be measured to have
an another kind. Maki et al. \cite{Maki} proposed
$\nu_{e}-\nu_{\mu}$ mixing as well as virtual $\nu_{e}-\nu_{\mu}$
transmutation. The flavor eigen states of neutrinos are the
coherent mixtures of their mass eigen states and thus the concept
of neutrino mass is evolved in the perspective of neutrino
oscillation. This type of neutrino oscillation is considered to be
the vacuum oscillation which can successfully explain the
`atmospheric neutrino anomaly'; but the vacuum oscillation fails
to explain the solar neutrino fluxes observed from the various
experimental data. Therefore, the matter effect must be taken into
account while studying the solar neutrino fluxes as the resonant
oscillation of neutrinos occurs in the solar matter. In 1978
Wolfenstein \cite{Wolfenstein} introduced the concept of
neutrino oscillation in presence of matter, but unfortunately he
used wrong sign of matter profile. Later, this matter effect on
neutrino oscillation was developed by Mikheev and Smirnov
\cite{Smirnov} to introduce the resonance phenomenon in neutrino
oscillation. Thus the influence of background matter on neutrino
oscillation is also known as MSW effect. In explaining the solar
neutrino data the MSW effect must be taken into account. In matter
background the electron type of neutrinos interact with electrons
in the matter, while the interaction effect of the muon (or tau)
neutrinos to the muons (or taus) is ignorable in the energy range
of solar neutrinos. As a result the nature of oscillation of
electron neutrino is quite different from that of other two
flavors and that is the key factor for introducing the MSW effect.
Eventually, the MSW  effect modifies the neutrino oscillation
phenomenon. Now the question may arise whether such effect can
generate the flavor transitions even in the framework of Standard
Model.\\\indent Wolfenstein claimed to show that if all neutrinos
are massless it is possible to have oscillations occur when
neutrinos pass through the matter. We shall examine whether such
demand has any justification. It is to be remembered that the
existence of neutrino mass is an ad-hoc assumption adopted to
explain the neutrino oscillation phenomenon. In other words if the
flavor transitions occur the neutrinos must have the non-zero
mass. But does the converse hold, i.e., can an effective mass
generate the neutrino oscillation? Which factor is then
responsible for the oscillation phenomenon? We must address such
questions arose. But at first for simplicity one can assume the
neutrino oscillation occurs only between two flavors: electron
neutrino and $l$ neutrino (which is a mixture of $\mu$ and $\tau$
neutrinos), whereas the initial beam consists of purely electron
neutrino flavors. In the section-2 we shall examine the
feasibility of matter induced oscillation in absence of any
neutrino rest mass. Mikheev and Smirnov could realized that the
presence of non-uniform background matter would produce the
density dependence effective mass. We shall verify whether such
effective mass can generate the neutrino oscillation in the
framework of standard model. In the section-3 we shall introduce
Stodolsky type of equation representing the dynamics of the
neutrino oscillation and study what information it can give about
the flavor transitions with zero rest mass of neutrinos, i.e., in
the framework of standard model.
\section{MSW effect in the framework of standard model}
In the scenario of flavor transition it is known that the flavor
eigen states are the superpositions of the mass eigen states with
the vacuum mixing angle $\theta_{0}$ giving the information about
the nature of mixing. Such mixing angle plays an important role in
the neutrino oscillations in vacuum. The maximum mixing takes
place for $\theta_{0}=\frac{\pi}{4}$. If the vacuum mixing angle
is 0 or $\frac{\pi}{2}$ there will be no mixing at all. Now let us
consider the evolution equation for the vacuum oscillation given
by
$$i\partial_{t}\left(%
\begin{array}{c}
  \nu_{e} \\
  \nu_{l} \\
\end{array}%
\right)=[E+\frac{m_{1}^{2}+m_{2}^{2}}{4E}\\+\frac{\triangle
m^{2}}{4E}
\left(%
\begin{array}{cc}
  -\cos 2\theta_{0} & \sin 2\theta_{0}\\
  \sin 2\theta_{0} &  \cos 2\theta_{0}\\
\end{array}%
\right)]
\left(%
\begin{array}{c}
  \nu_{e} \\
  \nu_{l} \\
\end{array}%
\right)\eqno{(2.1)}$$ The above equation is obtained in
approximating the three flavor case. It has already been stated
that the $\nu_{l}$ is taken as the superposition of $\nu_{\mu}$
and $\nu_{\tau}$. The approximations are valid because the mixing
angle $\theta_{13}$ is very small and because two of the mass
eigen states are very close to each other compared to the third
one. It is now quite easy to calculate corresponding transition
probability. We have taken the initial beam consists of electron
neutrinos and therefore the conversion probability becomes
$$P_{el}=|\nu_{l}|^{2}=\sin^{2}2\theta_{0}\sin^{2}(\frac{\triangle
m^{2}t}{4E})\eqno{(2.2)}$$ It is now observed that for
$\theta_{0}=0$ or $\frac{\pi}{2}$ there is no flavor transition
even if the neutrino has the non-zero mass. In that case each of
the flavor eigen states becomes either of mass eigen states. It is
worth noting that according to equation (2.2) the flavor
transition does not occur too if the neutrino has the zero mass.
It is quite clear that the vacuum oscillation depends on neutrino
mass as well as mixing angle and therefore, in vacuum the
neutrinos can oscillate only when they have the non-zero mass
along with the suitable vacuum mixing angle lying between 0 to
$\frac{\pi}{2}$.\\\indent Let us consider now MSW effect in which
both the eigen states and the eigen values, and consequently, the
effective mixing angle depend on matter density. In that case the
evolution equation is modified. The effective MSW Hamiltonian
takes the form
$$H_{f}=\left(%
\begin{array}{cc}
  -\triangle m^{2}\cos 2\theta_{0}+A & \triangle m^{2}\sin 2\theta_{0}\\
  \triangle m^{2}\sin 2\theta_{0} &  \triangle m^{2}\cos 2\theta_{0}\\
\end{array}%
\right)\eqno{(2.3a)}$$ where,
$$A=2E\sqrt{2}G_{F}n_{e}\eqno{(2.3b)}$$ The corresponding eigen
values of this matrix are
$$\widetilde{m}_{1,2}^{2}=\frac{A}{2}\mp\frac{\sqrt{(\triangle m^{2}\cos2\theta_{0}-A)^{2}
+(\triangle m^{2}\sin2\theta_{0})^{2}}}{2}\eqno{(2.4a)}$$ The
effective mass squared difference in presence of matter takes the
form
$$\triangle\widetilde{m}^{2}=\widetilde{m}_{2}^{2}-\widetilde{m}_{1}^{2}=\frac{\sqrt{(\triangle
m^{2}\cos2\theta_{0}-A)^{2} +(\triangle
m^{2}\sin2\theta_{0})^{2}}}{2}\eqno{(2.4b)}$$ We know that in the
framework of the standard model $m_{1}=m_{2}=0$ and hence
$\triangle m^{2}=0$. But in the matter induced oscillation (2.4a)
and (2.4b) imply $\triangle\widetilde{m}^{2}=A$. It means a
non-zero effective mass, proportional to the electron number
density, is evolved in presence of matter. Now we shall verify
whether such mass is capable of generating any oscillation. The
evolution equation for the matter induced oscillation is given by
$$i\partial_{t}\left(%
\begin{array}{c}
  \nu_{e} \\
  \nu_{l} \\
\end{array}%
\right)=[E+\frac{m_{1}^{2}+m_{2}^{2}+A}{4E}\\+\frac{1}{4E}\left(%
\begin{array}{cc}
  -\triangle m^{2}\cos 2\theta_{0} & \triangle m^{2}\sin 2\theta_{0}\\
  \triangle m^{2}\sin 2\theta_{0} &  \triangle m^{2}\cos 2\theta_{0}\\
\end{array}%
\right)+\frac{A}{4E}\left(%
\begin{array}{cc}
  1 & 0\\
  0 &  -1\\
\end{array}%
\right)]
\left(%
\begin{array}{c}
  \nu_{e} \\
  \nu_{l} \\
\end{array}%
\right)\eqno{(2.5)}$$ In the framework of standard model such
equation is reduced to
$$i\partial_{t}\left(%
\begin{array}{c}
  \nu_{e} \\
  \nu_{l} \\
\end{array}%
\right)=[E+\frac{A}{4E}+\frac{A}{4E}
\left(%
\begin{array}{cc}
  1 & 0\\
  0 &  -1\\
\end{array}%
\right)]
\left(%
\begin{array}{c}
  \nu_{e} \\
  \nu_{l} \\
\end{array}%
\right)\eqno{(2.6)}$$ We now compare the equation (2.6) with the
evolution equation of vacuum oscillation given by (2.1). It seems
to us that (2.6) represents the neutrino oscillation in which one
of the eigen state $\nu_{1}$ is massless and other one $\nu_{2}$
has an effective mass proportional to the electron number density.
It also shows that the corresponding effective mixing angle
becomes $\frac{\pi}{2}$ and therefore, the conversion probability
is zero, resulting there will be no oscillation at all. Due to
such mixing angle the electron flavor coincides with the heavy
eigen state i.e., $\nu_{e}\sim\nu_{2}$; as a result the electron
neutrino will gain the effective mass $2E\sqrt{2}G_{F}n_{e}$, but
fails to mix with $\nu_{l}$ that remains massless as per the
standard model consideration. Thus the flavor transition may not
be possible even if the electron neutrino gains the non-zero mass.
We shall see the situation from other angle.
\section{Pendulum in flavor space}
Let us now consider again the evolution equation of the vacuum
oscillation, i.e., the equation (2.1). It can be expressed as
$$i\partial_{t}\left(%
\begin{array}{c}
  \nu_{e} \\
  \nu_{l} \\
\end{array}%
\right)=[E+\frac{M^{2}}{2E}]
\left(%
\begin{array}{c}
  \nu_{e} \\
  \nu_{l} \\
\end{array}%
\right)\eqno{(3.1)}$$ where, the mass matrix $M^{2}$ is
represented by
$$M^{2}=\frac{m_{1}^{2}+m_{2}^{2}}{2}+\frac{\triangle m^{2}}{2}
\mathbf{B}.\mathbf{\sigma}\eqno{(3.2)}$$ where,
$\mathbf{B}=(\sin2\theta_{0},0,-\cos2\theta_{0})$ and
$\mathbf{\sigma}=(\sigma_{1},\sigma_{2},\sigma_{3})$ with
$\sigma_{i}$ ($i=1,2,3$) being the Pauli matrices. This
$\mathbf{B}$ plays an important role which will be discussed
later.\vspace{0.2cm}\\\indent A density matrix is defined by
$$\rho=\left(%
\begin{array}{c}
  \nu_{e}^{*} \\
  \nu_{l}^{*} \\
\end{array}%
\right)
\left(%
\begin{array}{cc}
  \nu_{e} & \nu_{l} \\
\end{array}%
\right)\eqno{(3.3)}$$ and now it is possible to express the
evolution equation of the vacuum oscillation in terms of density
matrix in the following form.
$$i\partial_{t}\rho=[M^{2},\rho]\eqno{(3.4)}$$
The corresponding density matrix can also be represented by
$$\rho=\frac{1}{2}(1+\mathbf{P}.\mathbf{\sigma})\eqno{(3.5)}$$
This $\mathbf{P}$ plays a crucial role to describe the dynamics of
the neutrino oscillation. The $z$-th component of $\mathbf{P}$ is
related to the transition probability as follows.
$$|\nu_{e}|^{2}=\frac{1}{2}(1+P_{z})\eqno{(3.5a)}$$
$$|\nu_{l}|^{2}=\frac{1}{2}(1-P_{z})\eqno{(3.5b)}$$
The other two components of $\mathbf{P}$ give the information
about the phase factors. $P=1$ stands for the perfect coherent
mixture of neutrino gas, but in reality $P<1$. Now using the
expressions of $M^{2}$ and $\rho$ from the equations (3.2) and
(3.3) respectively it can be deduced from the equation (3.4) as
$$\partial_{t}\mathbf{P}=\omega(\mathbf{B}\times\mathbf{P})\eqno{(3.6)}$$
where, $$\omega=\frac{\triangle m^{2}}{2E}\eqno{(3.6a)}$$ This is
analogous to a equation of a spherical pendulum in flavor space in
which $\mathbf{P}$ precesses round $\mathbf{B}$ at an angle
$2\theta_{0}$ with a frequency $\omega$. The representation of the
vacuum oscillation phenomenon in this form was first considered by
Stodolsky \cite{Stodolsky} and he deduced the equation (3.6). In
this case $\mathbf{P}$ stands for the polarization vector which is
very much similar to the polarization vector of light. It is now
clear that $\mathbf{P}=(0,0,1)$ represents the electron flavor and
$\mathbf{P}=(0,0,-1)$ stands for the $x$ flavor. Consequently,
$\mathbf{z}$ represents the flavor direction, whereas $\mathbf{B}$
stands as the mass direction. Such $\mathbf{B}$ is also called the
external magnetic field as the picture is very much analogous to
the spin-precession scenario of the atomic model, of course it has
no relation with magnetic field in the real sense. That pendular
model was frequently used to discuss the collective oscillation of
the supernova neutrinos \cite{Raffelt1, Raffelt2, Duan1, Duan2,
Duan3, Raffelt3, Mirizzi}.\\\indent We now consider the matter
induced oscillation. The equation (2.5) shows the evolution
equation of the MSW effect. Starting with that equation and
following the same procedure as in the case of vacuum oscillation
one can reach to the Stodolsky type of equation as
$$\partial_{t}\mathbf{P}=\omega(\mathbf{B}\times\mathbf{P})+\lambda(\mathbf{z}\times\mathbf{P})\eqno{(3.7)}$$
where, $$\lambda=\frac{A}{2E}\eqno{(3.7a)}$$ It shows the
polarization vector does not precess around $\mathbf{B}$ but can
be considered to precess around a vector
$\mathbf{R}=\omega\mathbf{B}+\lambda\mathbf{z}$ with frequency
$R$. The angle between such vector and $\mathbf{z}$ axis is
nothing but the twice of the mixing angle due to the matter
induced oscillation. The frequency and the effective mixing angle
are calculated as
$$R=\frac{\triangle \widetilde{m}^{2}}{2E}=[(\lambda-\omega\cos2\theta_{0})^{2}+\omega^{2}\sin^{2}2\theta_{0}]
^{\frac{1}{2}}\eqno{(3.8a)}$$ and
$$\cos2\widetilde{\theta}=\frac{\lambda-\omega\cos2\theta_{0}}{[(\lambda-\omega\cos2\theta_{0})^{2}+
\omega^{2}\sin^{2}2\theta_{0}]^{\frac{1}{2}}}\eqno{(3.8b)}$$ If we
watch the equation (3.7) we see the second part of that equation
seems to remain non-zero even while the vacuum frequency is taken
to be zero. Therefore, since the initial choice of $\mathbf{P}$ is
arbitrary it seems that the precession can takes place even
when $\omega=0$. Does it imply that oscillations occur even if the neutrinos are massless? Eventually, the answer would be negative. $\mathbf{z}$ represents the flavor directions, represented by $(0,0,1)$. In case, when $\omega=0$ the equation (3.7) yields $\partial_{t}P_{z}=0$ and thus $P_{z}=$ constant $=1$. Therefore, the survival probability remains 1, which must interpret that there is no oscillation at all.
\section{Discussion:}
From the Stodolsky type of equation it may be wrongly understood
that in the framework of standard model, i.e., if one can put
$\omega=0$ by hand the neutrino oscillation is still possible for
arbitrary polarization vector. But if we study the Stodolsky type
of equation very carefully it states the polarization vector
precesses around the negative mass direction (i.e., $-\mathbf{B}$)
in such a way that it must coincides, at a particular stage of its
motion, with the flavor direction (here it is $z$-axis). In case
of vacuum oscillation mass direction always makes a constant angle
with the flavor direction and therefore, polarization vector
precesses with that particular angle. In the matter induced
oscillation the angle between effective mass direction and flavor
direction is not constant at all, rather it depends on the matter
density. Consequently the neutrino oscillation depends not only
upon the mass, but it also depends on the mixing angle. In
presence of matter both of them are proportional to the electron
number density as well as the mass of the neutrinos. But in the
framework of standard model when the neutrino is massless in
vacuum the presence of matter yields an effective neutrino mass,
but fails to generate any oscillation as the effective mixing
angle becomes $\frac{\pi}{2}$, resulting no mixing at all. The
zero neutrino mass is equivalent to the vanishing vacuum frequency
$\omega$ in the Stodolsky equation. In this case the presence of
matter results a non-zero high frequency $\lambda$ but the mass
direction remains aligned to the flavor direction. For a flavor
oscillation, whatever small it may be, it is mandatory that the
mass direction gets separated from the flavor direction. As it is
not possible for $\omega=0$, i.e., for zero mass of either of the
mass eigen state no flavor oscillation can take place, even for a
high value of effective mass. Thus our analytical study shows that
the neutrino oscillation phenomena cannot be explained, even in
presence of matter if neutrinos are considered as massless.
Although the standard model is very much successful to explain
many low as well as high energy phenomena in the particle physics
but within the framework of this model it is not possible to
realize the neutrino oscillation phenomenon. Presently, there are
a number of extended models which can successfully explain the
generation of neutrino mass and henceforth the neutrino
oscillation.

\end{document}